\documentclass[twocolumn,prl,showpacs]{revtex4}
\usepackage{graphicx}
\usepackage{times}
\usepackage{amsmath,amssymb}

\newcommand{\eps}{\varepsilon}

\begin{document}

\title{Accessing the N\'eel phase of ultracold fermionic atoms in a simple-cubic optical lattice}

\author{C. J. M. Mathy}
\affiliation{Department of Physics, Princeton University,
Princeton, New Jersey 08544}
\author{David A. Huse}
\affiliation{Department of Physics, Princeton University,
Princeton, New Jersey 08544}

\date{\today}

\begin{abstract}

We examine the phase diagram of a simple-cubic optical lattice half-filled by two species of fermionic atoms %of mass $m$ %that
with a repulsive $s$-wave contact interaction.  We use the Hartree approximation in the regime of weaker interactions,
and a Hubbard model approximation in the Mott insulating phase. %we ask what conditions maximize the N\'eel
The regime where the N\'eel phase of this system is likely to be most accessible to experiments is
at intermediate lattice strengths and interactions, and our two approximations agree fairly well in this regime.
We discuss the various issues that may determine where in this
phase diagram the N\'eel phase is first produced and detected experimentally,
and analyze the quantum phases that occur in the intermediate lattice strength regime.
%temperature $T_N$ in the Mott insulator at density one atom
%per site, with equal numbers of the two species.  This maximum, $k_BT_N^{(\rm max)}\cong 0.15\,
%\hbar^2/(md^2)$, occurs near the edge of the regime where the
%system is well-approximated by the usual Hubbard model. The
%correction to the Hubbard-model approximation that produces a ``direct'' ferromagnetic interaction between atoms in
%nearest-neighbor Wannier orbitals is the leading term that limits how high $T_N$ can be made.
\end{abstract}

% 37.10.Jk atoms in optical lattices
% 71.10.Ca Electron gas, Fermi gas
% 71.10.Fd Lattice fermion models (Hubbard model, etc)
% 71.10.Hf Non-fermi-liquid ground states, electron phase diagrams and % phase transitions in model systems
% 05.30.Pr   Fractional statistics systems (anyons, etc.)
% 73.43.Lp  Collective excitations (Hall effects)
% 03.65.Vf   Phases: geometric; dynamic or topological
% 03.67.Lx  Quantum computation
% 73.43.?f   Fractional quantum Hall effect

\maketitle

%%%%%%%%%%%%%%%%%%%%%%%%%%%%%%%%%%%%%%%%%%%%%%%%%%%
% Introduction ---

One of the next notable milestones in the production of new
strongly-correlated states of ultracold atoms is
expected to be the antiferromagnetic N\'eel phase of two hyperfine
species of fermionic atoms in an optical lattice \cite{duan,
bloch}. Important progress towards this goal includes the recent
realization of the Mott insulating phase with fermions \cite{jordens, schn}, and the
demonstration of controllable superexchange interactions in an
optical lattice, albeit with bosonic atoms \cite{trotsky}.
The interest in experimentally realizing this N\'eel phase is partly
because it is a natural next step in developing an ultracold-atom
quantum ``simulator'' of the fermionic Hubbard model with repulsive
interactions, which is one of the
prototypical models in strongly-correlated quantum condensed-matter
physics.  As we will argue, if the objective is to access a phase with antiferromagnetic order,
then it may be favorable to explore the regime of an optical lattice of intermediate depth,
where the effective Hamiltonian is more than just the standard one-band Hubbard model.
The low temperature phase diagram in this regime is worth exploring in its own right,
as a new, interesting, and hopefully experimentally accessible
quantum many-body system.

An outline of our paper is as follows: first we review the physics of two-species fermions 
in an optical lattice, with the interactions tuned by a nearby Feshbach resonance.
We then describe the Hartree approximation which captures the ground-state phase diagram, and use this calculation
to estimate the line in the phase diagram along which antiferromagnetic interactions are strongest.
For another approach to parts of the same phase diagram, we map the system onto a lattice Hubbard model, 
which is appropriate for a strong lattice potential, and study which terms dominate the
Hamiltonian as one weakens the lattice away from the strong lattice regime. 
We also estimate the N\'eel temperature in this strong-lattice approach, and show that it
is maximized along a line in the phase diagram that agrees well with the line of maximal exchange interactions
estimated using the Hartree approximation.  Finally, we discuss experimental issues such as
equilibration, Feshbach molecule formation, and the size of the Mott phase in a trap.

\section{The model}

We consider the model that experimentalists are focusing on, which consists
 of two hyperfine species of fermionic atoms of mass $m$ in a simple-cubic
optical lattice with lattice spacing $d$ and a single-atom optical potential of the standard form
\cite{bloch}:
\begin{equation}
V_1(x,y,z)=V_0(\sin^2\frac{\pi x}{d}+\sin^2\frac{\pi
y}{d}+\sin^2\frac{\pi z}{d}) ~.
\end{equation}
Atoms of opposite species interact repulsively with a contact ($s$-wave)
interaction, whose strength can be adjusted
by tuning the applied magnetic field relative to a nearby Feshbach resonance, and can be quantified by the
atom-atom scattering length $a_s>0$ in the absence of the lattice.
We call these two states ``up'' and ``down'' and treat them
as the two states of a spin-1/2 degree of freedom.  The $s$-wave
repulsive interaction is only between atoms of opposite spin.  To
lowest order in $(a_s/d)$, this 2-atom interaction
is the standard regularized contact potential \cite{Huang},
\begin{equation}
V_2(\vec r_{\uparrow}-\vec
r_{\downarrow})=\frac{4\pi\hbar^2a_s}{m}\delta(\vec
r_{\uparrow}-\vec r_{\downarrow}) \frac{\partial}{\partial r}r ~,
\end{equation}
where $r$ is the distance between the two atoms.
 We will work in the
thermodynamic limit of this model at half filling, and we are particularly interested in the Mott insulating phase.

This system is appealing because it behaves effectively like a one-band Hubbard model at large $V_0$,
with on-site repulsive interaction $U$ and nearest-neighbor hopping matrix element $-t$, so that experiments
may eventually resolve long-standing questions about the Hubbard model. However, as $V_0$ is lowered one eventually
reaches an intermediate lattice-strength regime where higher bands can no longer be neglected.  To address this portion of the phase diagram,
we use the self-consistent Hartree approximation, which includes admixtures of the
single-atom states in higher bands.

\section{Hartree approximation}

When considering the regularized contact potential in mean-field theory, there is no exchange term,
implying that the Hartree and Hartree-Fock
approximations are identical here.  The total effective potential ``seen''
by the atoms with $S_z=-\sigma$ in the Hartree approximation is thus
\begin{equation}
V^{(eff)}_{-\sigma}(\vec r)=\frac{4\pi\hbar^2a_s}{m}n_{\sigma}(\vec r)+V_1(\vec r) ~,
\end{equation}
where $n_{\sigma}(\vec r)$ is the number density of atoms with $S_z=\sigma$ at position
$\vec r$.

For each point in the ground-state phase diagram, specified by the lattice intensity $V_0$
and the interaction $a_s$, we solve the Hartree equations numerically by discretizing
them in momentum space, and iteratively achieving self-consistency.
We obtain up to 3 different low energy self-consistent Hartree many-body states
at density one atom per lattice site, and determine which state is of the lowest energy and thus is the Hartree ground state.
In the paramagnetic state we impose the constraints
$n_{\uparrow}(\vec r)=n_{\downarrow}(\vec r)=n_{\uparrow}(\vec r+\vec \delta)$ at all $\vec r$, where
$\vec \delta$ is any nearest neighbor vector of the simple cubic lattice.  This paramagnetic Hartree
state exists for all values of $a_s$ and $V_0$, since this constraint is preserved by the iterations.
In this state the lowest Hartree bands are always partially occupied, so there are Fermi surfaces.
For the N\'eel antiferromagnetic state we impose only the constraints
$n_{\uparrow}(\vec r)=n_{\downarrow}(\vec r+\vec \delta)$, thus permitting two-sublattice N\'eel order.
At weak enough lattice and interactions, the only self-consistent Hartree state is in fact a paramagnetic state,
i.e. it has no magnetic order. % so there is no N\'eel state.
When an ordered N\'eel state exists, it can be either Mott insulating, with the lowest Hartree bands full
and a Mott-Hubbard gap to the next bands, or the two lowest Hartree bands for each species can overlap in energy and each be
partially occupied, with % it can have
Fermi surfaces.  We also look at ferromagnetic states that have different densities of
$\uparrow$ and $\downarrow$ atoms but do not break the discrete translational symmetries of the
lattice.  Again, there is a portion of the phase diagram where there is no self-consistent
ferromagnetic state.  And when there is such a state, it can be either a band insulator or
have Fermi surface(s), and it can be either fully or partially spin-polarized.  In the latter
case (PPF in Fig. 1) there are 3 partially occupied Hartree bands.  In these candidate
Hartree ground states, the fermions occupy only the lowest Hartree bands, but in terms
of free noninteracting fermions, we have included states extending out to many Brillouin zones
and thus the Hartree states are admixtures of multiple bands of the noninteracting
system. Specifically, for the parameters in Fig. 1, a $20\times 20\times 20$ grid of momentum
points in each Brillouin zone and a $9\times 9\times 9$ grid of zones was enough to achieve convergence everywhere.

\begin{figure}
\includegraphics[width=0.5\textwidth]{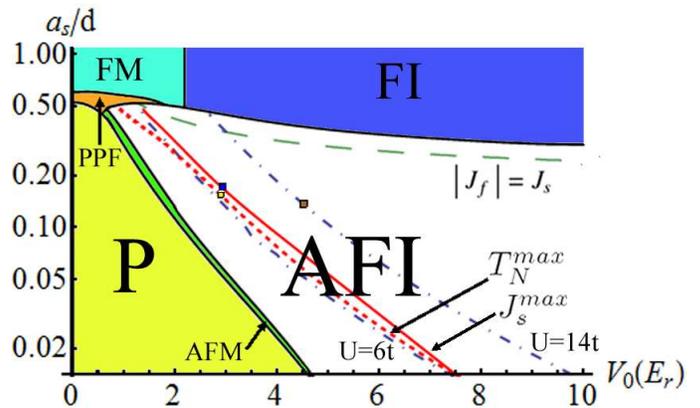}
\caption{Ground-state phase diagram for filling one fermion per lattice site %for fermions in a three-dimensional optical lattice.The ground state phase boundaries
in the Hartree approximation.
The phases shown are the antiferromagnetic Mott insulator (AFI); paramagnetic (P),
antiferromagnetic (AFM) and partly- (PPF) and fully-polarized (FM) ferromagnetic Fermi liquids; and the
ferromagnetic band insulator (FI).
%The line marked $|J_f|=J_s$ is our approximation to the
%ground-state phase boundary separating the antiferromagnetic phase
%at smaller $a_s$ from the ferromagnetic phase at larger $a_s$. The
%ferromagnetic phase is mostly a fully-polarized band insulator,
%but the bands overlap and it becomes a Fermi liquid at small %re is a small sliver of polarized Fermi liquid at small
%$V_0$, to the left of the line marked "FM to FI".
The solid line marked
$J_s^{max}$ indicates where the Hartree estimate of the effective exchange interaction $J$ as a function of the lattice strength $V_0$ is
maximized for each interaction $a_s$, and the dashed line near it shows where our estimate of the
N\'eel temperature $T_N$ gets maximized under the same prescription (see text). 
%The dashed line is obtained using a one-band approximation. This approximation starts breaking down along this line at about $9 E_r$. 
%, and at that point $J_f \simeq -J_s/4$ (see text).
The dash-dotted $U=14t$ line is near where the DMFT estimate of the entropy is maximized
at the N\'eel ordering temperature $T_N$ \cite{werner}. % and $T_N$ on this line is maximized at the square (see later in text).
The square on each line denotes the overall maximum of $J$ or $T_N$ along that line.
The lattice intensity $V_0$ is given in units of the recoil energy $E_r=(\pi\hbar)^2/(2md^2)$.}
%The dashed line marked $|J_f|=J_s$ is the estimate of the AFI/FI phase
%boundary based on adding the interaction perturbatively in
%the basis of the Wannier orbitals.}
%} %, at which $J_f \simeq -0.32 J_s$. %,  the entropy is maximized (according to QMC simulations\cite{staudt}). On both these lines, the highlighted point indicates where the N\'eel ordering temperature $T_N$ is globally maximized. \\
%Below the $J_f=J_s$ line, $J_s$ dominates over $J_f$, thus we expect an antiferromagnetic ground state below, and a ferromagnetic ground state above this line.\\
%The $t_0=t_I$ line signals when the interaction correction to the hopping %s become very significant, i.e. where the correction to nearest neighbour hopping is on the order of the hopping itself.\\
%becomes strong.  There is presumably also a Fermi
%liquid ground state on the lower left side of this phase diagram,
%but our approximations are not well-suited
%to estimating where this phase is.} % Finally, the "FM to FI" line demarcates a region
%below it in which the system is metallic, i.e. it becomes
%energetically favourable to form a hole and a spin flipped
%excitation over the fully polarized FM state.}
\label{fig:PhaseDiag}
\end{figure}

The resulting ground-state Hartree phase diagram of this system is presented in Fig. 1.
At any lattice strength, the paramagnetic Fermi liquid phase exists at weak enough interaction.
At strong interaction, the Hartree approximation always produces a ferromagnetic ground state, due to the classic Stoner instability. The ferromagnetic phase is a band insulator for $V_0\gtrsim 2.2 E_r$, with a band gap above the filled
lowest fully spin-polarized band. For $V_0 \lesssim 2.2 E_r$ the lowest two bands for the
majority-spin atoms overlap and there is instead a ferromagnetic
Fermi liquid.  We show this ferromagnetic part of the phase diagram for completeness,
but it is important to emphasize both that the Hartree approximation is not to be trusted at
such strong interactions, and that systems of ultracold fermionic atoms are likely to be highly
unstable to Feshbach molecule formation when brought this close to the Feshbach resonance.  Thus
we do not expect such a ferromagnetic phase to be experimentally accessible for these systems, even if it does
exist in a model that ignores the instability towards Feshbach molecules.  Finally, for lattices
above a certain minimum strength there is a N\'eel-ordered ground state at intermediate values of
the repulsive interaction.  This N\'eel state is a Mott insulator over most of the phase diagram, and
becomes an antiferromagnetic Fermi liquid over a small sliver of the phase diagram at weak lattice and weak interaction
between the paramagnetic and Mott insulating phases.  It is in this lower (smaller $a_s$) portion of the phase diagram
where we believe the Hartree approximation is qualitatively correct, producing the paramagnetic and
antiferromagnetic Fermi liquid phases and the Mott insulating N\'eel phase.

%there should also be a partially-polarized
%Fermi liquid ground state near the phase boundary to the N\'eel
%state. The transition from the fully-polarized band insulator to
%the partially-polarized ferromagnet occurs when the spin-polarized
%bands overlap, so the system can lower its energy by flipping
%spins.  A single spin flip makes a hole and a doubly-occupied site
%that are each moving freely within the fully-polarized background
%state.  At the level of approximation we have used in this paper,
%the hole moves freely with hopping $t_0$, so its lowest energy is
%$-6t_0$. The doubly-occupied state costs interaction energy
%$U_0+6U_{nn}$ and moves freely with effective hopping
%$t_2=t_0+2t_I$ because its motion is the hopping of the flipped
%spin between sites that are both occupied by unflipped spins.  The
%total energy of this particle-hole pair can be negative when
%$U_0+6U_{nn}<12(t_0+t_I)=6(t_0+t_2)$; this occurs below the line
%indicated in Fig. 1 as ``FM to FI'' (ferromagnetic metal to
%ferromagnetic insulator).

%well
%and the present
%approximations are not well suited to estimating the location of
%the corresponding phase transitions, so we leave that part of the
%phase diagram as ``{\it terra incognita}'' for now.
%[This stuff may belong later in the paper:]
%since the maximal $T_N$ occurs deep in the Mott insulating phase.
%where the Mott-Hubbard gap remains large compared to $t$.

To quantify the energy scale associated with N\'eel ordering in the antiferromagnetic phase, we (crudely) estimate the nearest-neighbor antiferromagnetic
exchange $J$ by taking $J/2$ to be the energy difference per bond between the
Hartree antiferromagnet and ferromagnet (at low $a_s$ and/or $V_0$, the ferromagnet has zero magnetization, so is really
the paramagnet).
For each given interaction $a_s$, we locate the lattice intensity $V_0$ where this Hartree
estimate of $J$ has its maximum, and indicate these maxima by the red (full) line marked $J_s^{max}$ in Fig. 1.
It is somewhere near this line that antiferromagnetic
interactions are strongest, and the N\'eel phase
survives to the highest temperature.
The overall global maximum
in this estimate of $J$ occurs at $a_s/d\cong 0.15$ and $V_0\cong 3E_r$, where $E_r=(\pi\hbar)^2/(2md^2)$ is the
``recoil energy''.  This point is marked with a small square on the full red line in Fig. 1.

\begin{figure}
\includegraphics[width=0.5\textwidth]{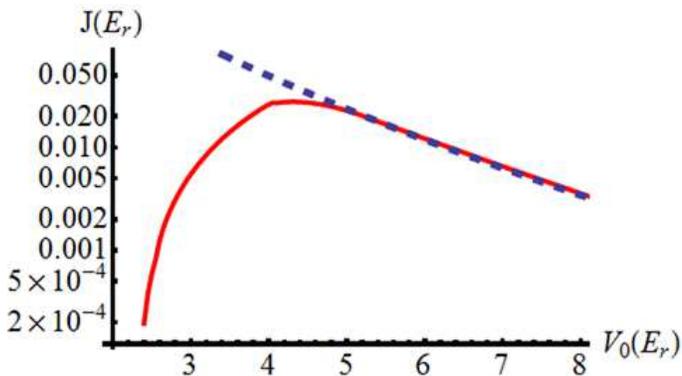}
\caption{
Plot of the Hartree estimate of the antiferromagnetic exchange coupling, $J$, as a function of $V_0$ for $a_s/d=0.08$,
compared with the estimate from the perturbative expansion at strong lattice. The red (full) line shows the Hartree estimate of $J$,
while the blue (dotted) line gives the perturbative estimate, discussed in the text.
They match very well at high lattice depth $V_0$.}
%Note that at around $V_0=4 E_r$, the Hartree energies of the paramagnetic and the ferromagnetic cross, thus signaling a transition from an SDW-like state to a state governed by local moments.}
\end{figure}

As an example, in Fig. 2 we show
our Hartree estimate of $J$ vs. $V_0$ at $a_s/d=0.08$, which is the largest interaction explored experimentally in the
study of ref. \cite{jordens}, although these experiments only reached temperatures far above those of the N\'eel phase. In Fig. 2 we also plot the estimate of $J$ from a strong lattice expansion which we  discuss in the next section, and it coincides well with the Hartree estimate for $V_0>5 E_r$.

The line in Fig. 1 where the Hartree estimate of $J$ is maximal not only indicates roughly where $T_N$ is maximized, but
it also occurs at the location of a crossover in the nature of the antiferromagnetic state.  In the regime
of larger $V_0$ and $a_s$, the system is a ``local moment'' magnet: almost every well of the optical lattice is
occupied by a single atom with a ``spin'', with very few wells either empty or multiply-occupied.  Here we
can make a Hartree ferromagnet as well as an antiferromagnet and compare their energies to obtain our estimate of $J$.
In fact we could make many other metastable magnetic Hartree states with ordering at essentially any momentum in the Brillouin zone.
Here $J$ is primarily a superexchange interaction $J\sim t^2/U$, and it increases as the lattice depth $V_0$ is reduced (see Fig. 2), since this increases
the hopping $t$ between wells and reduces the interaction $U$ between two unlike atoms in the same well.

In contrast, the regime of the antiferromagnetic phase that is at lower $V_0$ and $a_s$ near the paramagnet is not
a local-moment regime, but instead is a spin density wave (SDW).  Here one can make a magnetically-ordered Hartree state only
at momenta near those that cause substantial Fermi-surface nesting, which is initially only near the corners of the
first Brillouin zone.  In this regime we do not have a ferromagnetic Hartree state and our estimate of $J$ is obtained by subtracting
the energy of the N\'eel state from the paramagnet, so this $J$ should not be interpreted simply as
a spin-spin interaction.  We now understand why this estimate of $J$ increases with increasing $V_0$ (see Fig. 2), since this increases the
interaction that causes the magnetic ordering, and decreases the hopping that favors the paramagnetic Fermi liquid.  The
effective $J$ is maximized at the crossover between the SDW and local moment regimes.  This crossover shows up in the
metastable ferromagnetic Hartree state as two closely-spaced phase transitions between un-, partially-, and fully-polarized.

%The change of slope of $J$ at about $4 E_r$ can be understood by forgetting by comparing the paramagnet to the ferromagnet.
%The Hartree estimates of their energies cross at about $V_0=4 E_r$ and $a_s/d=0.08$ (there is a very thin PPF phase between P and F,
%so the system quickly becomes fully magnetized as one increases V). Since these parameters are sitting safely in the Mott phase,
%this phase transition is fictitious at zero temperature. However, if we were to disrupt AF order by increasing the temperature for example,
%this phase transition might be restored. At values of $V_0<4 E_r$, the phase would be gapless and its low energy excitations would be
%spin density waves. For $V_0>4 E_r$, the phase would be gapped, and the spins would each be localized on a lattice site and pointing in the same direction.
%Their hopping would be blocked by Pauli exclusion, and low energy excitations would be local spin flips. Since to obtain $J$ we compare the
%energy of the AF to that of P or F depending on which has lower energy, the crossover from the SDW to the local moment regime is mirrored in the estimate of $J$.

\section{Strong lattice expansion}

%To verify our results of the Hartree calculation, w
We now proceed to mapping our system onto a lattice 
Hubbard model, which is an appropriate description at deep lattice.
This mapping uses the
single-atom Wannier orbitals as the basis states
\cite{kohn,duan,werner}.  The standard one-band Hubbard model
includes only the lowest-energy Wannier orbital at each lattice
site as its Hilbert space, and neglects all terms in the Hamiltonian except for nearest-neighbor hopping and on-site
interaction between different hyperfine states in the lowest band.

The single-atom optical-lattice potential (1) that we consider is a separable potential, so the energy eigenstates of a
single atom in this potential can be chosen to be the product of
one-dimensional (1D) eigenstates along each %of the three
direction.  We solve for these 1D bands, and
use the gauge degree of freedom in the definition of Wannier states to minimize the spatial variance of their probability distributions.
Thus we obtain the normalized wavefunctions $w_n(x)$ of the maximally-localized 1D Wannier orbitals for each band $n$ ($n=0,1,2,..$) \cite{kohn}. %\footnote{These are the unique Wannier functions to be real, either symmetric or antisymmetric, and fall off exponentially with distance }.
The 3D Wannier orbitals are then
$\phi_{n_xn_yn_z}(x,y,z)=w_{n_x}(x)w_{n_y}(y)w_{n_z}(z)$.

Since our focus here is on the Mott insulator with density
one atom per lattice site, low temperature and fairly strong lattice $V_0$, the
atoms will primarily be in the lowest band, $n_x=n_y=n_z=0$.  The
nearest-neighbor hopping matrix element $t_0$ in this noninteracting single-atom band
is a strongly decreasing function of $V_0$, the lattice strength, behaving as
%$t_0\approx 4 \pi^{-1/2} E_r^{1/4} V_0^{3/4} e^{-2\sqrt{V_0/E_r}}$
$t_0\sim V_0^{3/4} e^{-2\sqrt{V_0/E_r}}$ for large $V_0$ \cite{werner}.
%One dimensionless small parameter that
%is important in what follows is the ratio of the amplitude of the Wannier orbital in
%$t_0/\eps_0$, where
The wavefunction $w_0(x)$ of the lowest Wannier
orbital at a given lattice site has its maximum
amplitude at $x=0$, the center of that well of the optical lattice,
%at that site,
while its amplitude changes sign and is of smaller magnitude %by a factor of order $t_0/\eps_0$
in the nearest-neighbor wells. %of the lattice.
%The ratio of these amplitudes is one small parameter that is important in the approximations we use below.
%We find $|w_0(x_{\rm min})/w_0(0)|\sim 3 t_0/\eps_0$, where
%$x_{\rm min}>d$ is the location of the absolute minimum of
%$w_0(x)$ and $\eps_0$ is
%the expectation value of the single-particle energy in a lowest 3D
%Wannier orbital.

The antiferromagnetic Mott insulator occurs in the regime where the on-site repulsive interaction $U$ is
stronger than the hopping $t$.  %We do not treat the regime
%of weak interaction where the system is a Fermi
%liquid that can have spin-density-wave order, since in this
%regime $T_N$ is very low or zero.
The expectation value of this interaction energy for two atoms of
opposite spin occupying the same lowest Wannier orbital is our
first estimate of the strength of the on-site interaction
$U_0n_{i\uparrow}n_{i\downarrow}$ %at site $i$ of
in the corresponding
one-band Hubbard model:
\begin{equation}
U_0=\frac{4\pi\hbar^2a_s}{m}[\int dxw_0^4(x)]^3 ~.
\end{equation}

In the Hubbard model, when adjacent sites $i$ and $j$ are each
singly-occupied by atoms with the same spin, then the hopping between those
two sites is Pauli-blocked.  When these adjacent sites are each
singly-occupied by opposite spins, then virtual hopping between
these sites, treated in second-order perturbation theory, allows
them to lower their energy and thus generates the antiferromagnetic
superexchange interaction $J_s(\vec S_i\cdot\vec S_j-\frac{1}{4})$
with $J_s=4t^2/U$.

The leading corrections to the Hubbard model approximation to our
system in the regime of interest are due to the
interactions between atoms of opposite spin occupying lowest
Wannier orbitals on nearest-neighbor sites $i$ and $j$.  %There are
%2 contributions:
By expanding fully the interaction term in the Wannier basis, and evaluating all the terms,
one can show that the most important term in limiting the N\'eel phase %how large $T_N$ can be made,
is the ``direct'' interaction \cite{trotsky}
\begin{equation}
U_{nn}=\frac{4\pi\hbar^2a_s}{m}[\int dxw_0^4(x)]^2\int
dyw_0^2(y)w_0^2(y+d)
\end{equation}
between atoms of opposite spin in adjacent orbitals.  This term is due to the overlap of the probability distributions of adjacent Wannier orbitals.  It
raises the energy of the N\'eel state, and thus produces a direct
ferromagnetic exchange interaction $J_f(\vec S_i\cdot\vec
S_j-\frac{1}{4})$ with $J_f=-2U_{nn}<0$ that partially cancels the
antiferromagnetic superexchange $J_s$ that occurs in the Hubbard
model.  It is primarily this ferromagnetic interaction that stops
and reverses the increase in the net antiferromagnetic interaction $J$ as one moves towards stronger
interaction and a weaker lattice while staying near the
values of the lattice strength $V_0$ that maximize $J$.  At the global maximum of $J$, indicated in Fig.
\ref{fig:PhaseDiag}, we find $J_f \simeq -J_s/4$.
Another (small) effect of $U_{nn}$ that we include is that it changes the effective on-site Coulomb interaction in the
antiferromagnetic Mott phase, leading to an effective interaction $U=U_0-6U_{nn}$.

For large enough $a_s$ this direct ferromagnetic
exchange is stronger than the superexchange and thus we have the
ground-state phase transition to the ferromagnetic phase, as
indicated in Fig. 1.   We see in Fig. 1 that the line where $|J_f|=J_s$ is
close to the line where the Hartree calculation also signals this phase transition.
If this transition is real, further-neighbor interactions might cause other
magnetically-ordered phases to occur near it in the phase diagram, 
but this is an issue that we do not explore here, since we expect that this
portion of the phase diagram will not be accessible to cold-atom
experiments, as discussed above.
  %But this
%regime seems unlikely be to experimentally accessible, so we do
%not explore it further here.  In addition, the approximations we
%are making may become quite inaccurate at these large values of
%$a_s/d$.

%The ``direct'' nearest-neighbor interaction (4) also slightly reduces the
%effective $U$ that enters in the superexchange interaction.  At
%this level of approximation our Hubbard model has
%interaction $U=U_0-6U_{nn}$, since it is the {\it change} of the
%interaction energy due to moving the atom that enters in the
%energy denominator in the superexchange process.

The next leading correction to the Hubbard model is an additional hopping term of the
same sign as $t_0$, sometimes called correlated hopping in the literature \cite{werner},
\begin{equation}
t_I=-\frac{4\pi\hbar^2a_s}{m}[\int dxw_0^4(x)]^2\int
dyw_0^3(y)w_0(y+d)
\end{equation}
%that is operative
when the two sites are each singly-occupied by
opposite spins. The effective hopping that enters in the
superexchange process at this level of approximation is thus
$t=t_0+t_I$.  More generally, the hopping matrix element for an atom
with $S_z=-\sigma$ hopping between nearest-neighbor sites $i$ and $j$
is $t=t_0+n_{\sigma}t_I$, where $n_{\sigma}$ here is the total number of
atoms with $S_z=\sigma$ on sites $i$ and $j$.

Summarizing, once we include these leading effects due to the
nearest-neighbor interaction, the effective Hamiltonian in the vicinity of
the antiferromagnetic ground state of the half-filled Mott insulator has hopping $t=t_0+t_I$, an
effective on-site interaction $U=U_0-6U_{nn}$, and an additional
ferromagnetic nearest-neighbor exchange interaction $J_f=-2U_{nn}$
when both sites are singly-occupied. [We note in passing that there are other condensed matter systems
which behave effectively like a Hubbard model with ferromagnetic interactions, such as $V_2O_3$ \cite{Kotliar}.]
 We can use this approximation to
produce an estimate of $U/t$ for each point in the Mott insulating regime
of our phase diagram, and we indicate in Fig. 1 the lines along which $U/t=6$ and $U/t=14$.

Now that we have our effective model, we use it to determine which values of the parameters will
maximize the robustness of the antiferromagnetic Mott phase. To that end, we start by looking at the Quantum Monte Carlo
simulations \cite{staudt} of the simple-cubic fermionic Hubbard model, which show that for a given $U$, the
critical temperature $T_N$ for the N\'eel phase is maximized near $U/t=6$.
From Fig. 1, we learn that the line along which $U=6t$, as estimated from the Hubbard-model approximation is
in reasonably good agreement with the location of the maximum $J$ as
estimated in the Hartree approximation. Thus we can hope to start from the QMC result for $T_N$, and study perturbative corrections to it
using the strong-lattice expansion, to obtain a reasonable approximation to $T_N$ four our model.

To estimate the N\'eel
temperature of our system we propose the following approximation:
For the Hubbard model without $J_f$, we have estimates of its
N\'eel temperature $T_N^{(H)}(t,U)$ from quantum Monte Carlo
simulations \cite{staudt}.  In the regime of interest, this N\'eel ordering is due to the
antiferromagnetic superexchange interaction $J_s=4t^2/U$ between
neighboring singly-occupied sites. When we include $J_f<0$ the magnetic interaction is reduced, and we will approximate the
resulting reduction of $T_N$ as being simply in proportion to the
reduction of the total nearest-neighbor exchange interaction:
\begin{equation}
T_N(V_0,a_s)\cong(1+\frac{J_f}{J_s})T_N^{(H)}(t,U) ~.
\end{equation}
This approximation is certainly accurate for large $U/t$, where the
system is well-approximated by the Heisenberg antiferromagnet.  By using this
approximation at the maximum of $T_N^{(H)}(t,U)$ near $U\cong 6t$, we are assuming that
the reduction there by roughly a factor of 2 of $T_N$ from its
Heisenberg value \cite{staudt} is mostly due to the dilution of
the Heisenberg antiferromagnet by nonmagnetic empty and
doubly-occupied sites and not due to a large change in the
superexchange interaction between singly-occupied sites.

Using eq. (7), we can maximize $T_N$ as a function of $V_0$ for each value of $a_s/d$.  The
resulting ``optimal'' values of $V_0$ are indicated by the dashed red line in Fig. 1.
The line of maximal $T_N$ we obtain this way is quite close to the full red line along which the Hartree estimate of $J$ is maximized.
The global maximum of this estimate of $T_N$ (marked by a small square on the line in Fig. 1) also coincides well with the global maximum of the Hartree $J$.
Thus we see that in this regime these two approximations agree rather well, suggesting that these estimates are indeed
to be trusted.
 %The highest $T_N$ occurs near $a_s/d=0.15$, but the
%system at this value of $a_s$ is may be too close to the
%Feshbach resonance and thus not stable against formation of
%molecules.  The highest achievable $T_N$ thus may be
%somewhere along this line at a lower value of $a_s$ and thus a
%stronger lattice $V_0$. We note that a recent experiment \cite{jordens} has
%studied $a_s/d \simeq 0.08$ for $^{40}K$, albeit at a temperature well above $T_N$,
%without noting any strong instability towards molecule formation.
We also show the line along
which $U=14t$; this is near where DMFT calculation show the critical entropy
$S(T_N)$ to be maximal \cite{werner}. %so if the system can be
%adjusted adiabatically this is near where the N\'eel phase is most
%accessible.
Along this $U=14t$ line the maximum in $T_N$ also occurs near $a_s/d=0.15$.

\begin{figure}
\includegraphics[width=0.5\textwidth]{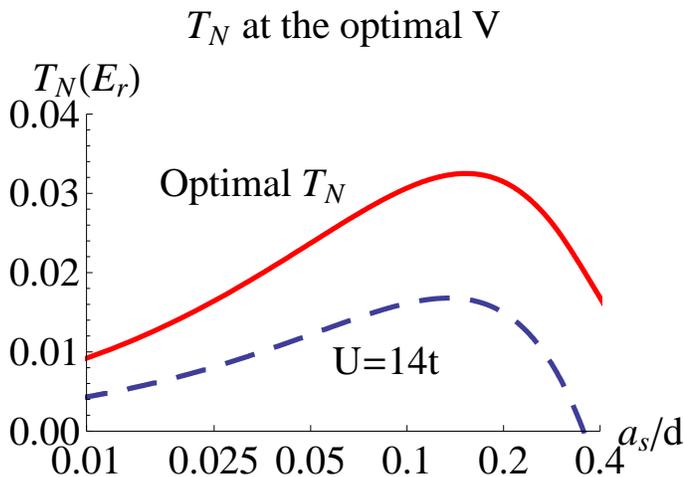}
\caption{Solid line: our estimates of the maximal N\'eel temperature, $T_N$,
as a function of $a_s/d$. For each value of $a_s$, $T_N$ is
maximized by varying the lattice depth $V_0$.  Dashed line: we also plot $T_N$
at the line $U=14t$, which is near where the critical entropy is
maximized in DMFT \cite{werner}.}
\end{figure}

In Fig. 3 we show $T_N$ as a function of
$a_s/d$ at the value of $V_0$ that maximizes our estimate of
$T_N$, as well as at the value of $V_0$ that gives $U=14t$ and
thus is near the maximum of $S(T_N)$ as estimated in DMFT \cite{werner}.  Note that in Fig. 3 the horizontal
scale for $a_s/d$ is logarithmic, so $T_N$ drops rather weakly as $a_s$ is reduced.
This means that if the optimal value of $a_s$ can not be reached due to Feshbach
molecule formation, this may not ``cost'' very much in terms of the suppression of $T_N$.
 On the other hand, increasing $U/t$  can produce a substantial decrease in $T_N$: it drops by about a factor of two
between its maximum at $U\cong 7t$ and $U=14t$.  For $^6$Li and $d \cong 500$ nm, the overall maximum of
$T_N$ is a temperature of about 40 nK.

The strong lattice approximations we have used here are those appropriate for the antiferromagnetic Mott insulator,
and are based on the inequalities on energy scales $\eps_0>U>t$, where $\eps_0$ is the expectation value of the single-particle energy in a lowest Wannier orbital.
By calculating higher order terms, we can check the reliability of the estimates coming from this strong lattice expansion.
We do find that this expansion is beginning to break
down in the vicinity of the parameter values that maximize $J$.  Thus, although we expect these approximations to give reasonably reliable rough estimates of the %location of this optimum and the
maximal values of $T_N$, there are many higher-order effects that we are ignoring that may alter these estimates (our calculations suggest on the $\sim 20\%$ level).
At the maximum of $T_N$, $|J_f|$ is about 25\% of $J_s$.  The correction to $J_s$ due to $t_I$ is also of roughly this size, but its dependence on $a_s$ is much weaker, which is why $J_f$ is the important actor in setting the maximum in $T_N$.

\begin{figure}[htb]
\includegraphics[width=0.45\textwidth]{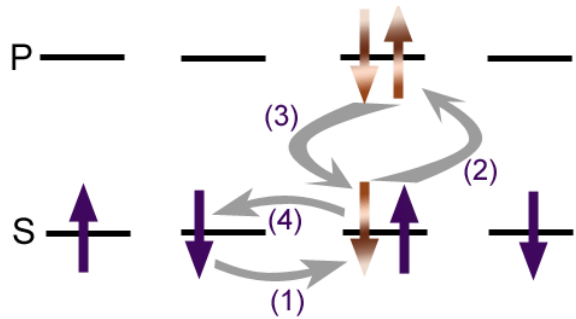}
\caption{The strongest higher-order process contributing to the
energy of the antiferromagnetic Mott insulator at the maxima of
$T_N$ shown in Fig. \ref{fig:PhaseDiag}.  It consists of (1) a
nearest-neighbor hop in the lowest (S) band, (2) an on-site ``pair
hopping'' of both atoms up to the next (P) band, (3) on-site
pair hopping back to the S band, and (4) a nearest-neighbor hop
back to the original configuration.  At both maxima of $T_N$, this
process corrects $J_s$ by about 10\%.
This indicates that mixing with higher bands is significant,
which we don't capture with our strong lattice expansion, but do capture with our Hartree calculation.
}
\label{fig:VirtualHop6}
\end{figure}

We have analyzed in perturbation theory many corrections beyond those included above.  We find that at %and it turns out that the largest correction appears at fourth order in perturbation theory, see Fig. \ref{fig:PhaseDiag}). At
the maximum of $T_N$ (both the global maximum and the maximum
along the $U=14t$ line) the strongest next correction is the
fourth-order process illustrated in Fig. \ref{fig:VirtualHop6}; it
alters the effective $J$ by about 10\%.

\begin{table}[h]
\caption{The values of the various energies at the two $T_N$ maxima. The ``$J_s$ correction'' corresponds to the process detailed in Fig. \ref{fig:VirtualHop6}.}
%\begin{figure}
\begin{tabular}{c | c | c}
        &   Global maximum      & Maximum\\
        &   of $T_N$            & of $T_N$ at $U=14t$ \\ \hline
$\eps_0\, (E_r)$ &    4.2      &    5.5    \\
$U_0\,(E_r)$    &  0.9  &   1.3      \\
$t_0\,(E_r)$    &   0.11    &   0.07 \\
$t_I\, (E_r)$   &   0.02    &   0.02 \\
$J_s\, (E_r)$       &   0.08    &   0.03 \\
$J_f\, (E_r)$       &   -0.02   &   -0.008 \\
$J_s$ correction $(E_r)$  & 0.007   & 0.004 \\
\end{tabular}
%\label{fig:Tab}
%\end{figure}
\end{table}

%We have also calculated the ground-state phase diagram and the net %effective exchange $J$ of this system in the Hartree-Fock (HF)
%approximation.  A portion of the HF phase diagram is shown %in Fig. 1.  The location of the largest HF estimate of $J$, and thus
%we expect the largest $T_N$, agrees well with the treatment presented %above that is
%based on corrections to the Hubbard model within the antiferromagnetic %Mott insulator \cite{hf}.
%Although neither approximation treats the quantum spin fluctuations %accurately, HF does allow
%strong mixing between Wannier bands, so this agreement suggests that %band mixing
%is not important in setting what parameters maximize $T_N$.

We have seen that our two approaches give consistent results in the region of intermediate lattice strength.
We now briefly discuss possible improvements on what we report here. Since our perturbatively-based approximations may be breaking down near
this regime of interest where $J$ is maximized, it would be nice
to have a more systematic approach that can obtain more precise
and reliable estimates of the phase diagram in this regime.  For
example, quantum Monte Carlo simulations might be possible at this maximal $T_N$, although of course the famous fermionic
``minus signs'' may prevent this from being feasible in the near term. Another approach would be a systematic correction of the Hartree calculation.
%this
%process leads to a negative correction to the energy of the $AF$
%state equal to about $- 0.03 J_s$ per bond. Since this process
%involves on-site pair hopping into a higher band, we expect the
%largest correction to our calculations to be on-site band mixing.

Since our calculations are in the thermodynamic limit, we are assuming that finite-size effects do
not substantially shift the transition at $T_N$ or alter the ground-state phase diagram.
The experiments use clouds of well over $10^4$ atoms \cite{jordens},
so this should be true unless a trap of extreme aspect ratio
is used that reduces a linear dimension of the Mott domain to
well under $10d$.

\section{Experimental consequences}

We are now in a position to discuss the experimental consequences of the calculations above.
The assumption usually found in the literature is that experiments will be able to evolve the system
adiabatically, and that the natural variable is therefore the entropy, instead of the temperature.
We would like to point out some caveats to this.

The system must be cooled to low temperature $T$ and thus low
entropy $S$, and equilibrated at the point in the phase diagram that is
being measured.  This cooling may be done under some other conditions
(e.g., with the lattice turned off and other parameters optimized for cooling),
with the ``pre-cooled'' system then moved adiabatically to the
conditions of interest for measurement \cite{duan,werner,dare,koetsier,snoek,deleo}.
For this to work, the time scales
of the system must be such that this can be done without strongly violating
adiabaticity.  For the N\'eel phase of the Mott insulator, this will limit
how small the hopping energy $t$ and the antiferromagnetic superexchange interaction $J$ can be,
since the system must be able to remain near equilibrium, adiabatically rearranging the atoms so that there
is one atom per lattice site and these atoms are antiferromagnetically correlated.
Alternatively, the system
might be actively cooled under the conditions of measurement, but this again
requires the system to be able to equilibrate under those conditions.  Thus, either
way, this constraint limits equilibrium access to the strong-lattice
portion of the phase diagram, where the exchange $J$ and/or hopping $t$ are too slow to allow equilibration
and adiabaticity.  Thus the N\'eel phase is going to be most accessible to experiment in some regime
of intermediate lattice strength $V_0$.  These considerations may also limit how large the interaction can be made, since
strong repulsion $a_s$ suppresses the superexchange rates and thus can limit spin
equilibration in the Mott insulating phase. %However, from Fig. 3 we learn that $T_N$ depends weakly on $a_s$ (since the $a_s$ scale is logarithmic),
%meaning that experimental limitations on $a_s$ will not drastically restrict $T_N$.

Another constraining issue is that as one approaches the Feshbach resonance
from the repulsive side, two atoms will interact repulsively only as long as they
scatter in a way that is orthogonal to the molecular bound states.  Thus the
atoms must remain metastable against forming Feshbach molecules.  In the
absence of the lattice, the rate of molecule formation grows as $\sim a_s^6$
as the Feshbach resonance is approached \cite{petrov}.  This will
limit how large the interaction $a_s$ can be made, in a way that is presumably
less of a constraint as the lattice is made stronger so that it keeps the
atoms apart.

As we mentioned earlier, the DMFT estimate of the maximum of $S(T_N)$ is near $U=14t$ \cite{werner}. In fact, the maximum in $S(T_N)$ is rather weak, being
only a little higher than that of the Heisenberg limit $U\gg t$.  Thus one might more
usefully say that the critical entropy is nearly maximized for any $U>12t$. As we can infer from Fig. 1, $U>12t$ occurs at relatively large $a_s$ and $V_0$,
and the issues raised above may force experiments away from the maximal entropy line,
and towards the line where the exchange interactions and $T_N$ are maximized.

So far we have assumed an homogeneous system,
but the trapping potential is actually nonuniform, which leads to inhomogeneity of the local equilibrium state
in the trap. The spatial size of the region occupied by the antiferromagnetic Mott phase will increase with increasing $U/t$,
since $U$ increases the Mott ``charge'' gap.  This effect means larger $U/t$ should favor detection %accuracy of detection methods
of the N\'eel phase; of course the optimal $U/t$ will be some compromise between this and the other issues discussed above.  % may favor large $U$.
Using our Hartree calculation to obtain the charge gap in the Mott phase, we find within the local-density approximation (LDA)
that at the point in the phase diagram where $J$ is maximized, the Mott phase should occupy about half of the linear size of the trap.
This is encouraging, but the Hartree approximation likely overestimates the Mott gap,
as the true gap should be renormalized downwards by spin fluctuations.

Finally, we come to novel experimental predictions that result from our calculations.
The Hartree calculation predicts that there is interesting and as yet unexplored physics at low to intermediate lattice strength, and weak coupling.
We mentioned already that the effective model at around optimal values of $V_0$ and $a_s$ is that of a Hubbard model with ferromagnetic correlations,
which is interesting in its own right. The different phase boundaries that we uncovered may also be worth exploring.
For example, the quantum
phase transitions between the Mott and metallic N\'eel phases and the paramagnet
occur in parameter regimes that are quite accessible to the
experiments, although it may not be possible to see their effects at
accessible temperatures, since $T_N$ decreases strongly as this weak-coupling
regime is approached.
%Since we restricted ourself to certain mean field ground states, there may be other magnetically-ordered phases.
%Such phases would be most likely close to the phase boundaries we uncovered, such as the phase boundary between the Mott state and the Ferromagnetic insulator where
%$|J_f|=J_s$ so the nearest-neighbor exchange cancels and further-neighbor interactions may become important.

\section{Conclusion}
We have shown that to maximize antiferromagnetic interactions for fermionic atoms in an optical
lattice one must explore the regime of intermediate lattice depths, where the system has significant
%enter regions of parameter space where the Hamiltonian is more intricate than a
deviations from the standard one-band Hubbard model.  We have found that the
nearest-neighbor direct ferromagnetic exchange is the most
important correction to the Hubbard model that limits the maximal exchange $J$,
and therefore the maximal N\'eel temperature $T_N$.

There are also higher-order corrections to the Hubbard model:
virtual hopping into higher bands and other higher-order processes.
The relative contribution of the higher-order corrections in the
vicinity of the optimal $J$ drops exponentially as one goes to
smaller interaction $a_s$ and thus a larger $V_0$.
Thus our perturbative calculation should yield accurate results in the large $V_0$ (strong lattice) regime.

We included a subset of all higher-order corrections via a Hartree calculation, and used it to find an estimate of the line
where the exchange $J$ is maximized. This line coincides well with the line where our estimate of $T_N$ is maximized,
obtained by using quantum Monte Carlo results and the strong lattice expansion.

For quantitatively more accurate results in the
intermediate lattice depth regime, one needs to resort to more systematic
quantum calculations.  Of course this is a system of many
fermions, so it is not clear whether this regime can
be accurately treated in some form of quantum Monte Carlo
simulations.

Experiments on these systems still need to reduce the temperature by a substantial factor before they are able to access
magnetically ordered phases of fermions in optical lattices.
Once they bridge this gap, our results above suggest that the
N\'eel phase will be most accessible %first magnetic phases they will be able to access will likely be
in the intermediate lattice strength regime.
If that is the case, then this ``quantum simulator'' should be able teach us about
more than just the standard one-band Hubbard model.

%s with ferromagnetic correlations, and be able to
%the quantum phase transitions in this region of parameter space.

%In the limit of no interaction, $a_s=0$, the ground state of this
%system is a Fermi gas with the lowest band half-full, so it is not
%Neel ordered.

%%%%%%%%%%%%%%%%%%%%%%%%%%%%%%%%%%%%%%%%%%%%%%%%%%%
% Acknowledgments ---
\section{Acknowledgements}
We thank Randy Hulet for many discussions, and Meera Parish for helpful suggestions. We also thank Roberto Car and Xifan Wu for advice on setting up the Hartree calculation. Part of this paper has been posted earlier as arXiv:0805.1507.  This work was
supported under ARO Award W911NF-07-1-0464 with funds from the
DARPA OLE Program.

%%%%%%%%%%%%%%%%%%%%%%%%%%%%%%%%%%%%%%%%%%%%%%%%%%%

% References ---

\end{document}